\documentclass[aps,twocolumn,showpacs]{revtex4}
\usepackage{epsfig}
\usepackage{amsbsy}
\usepackage{amsmath}
\usepackage{amsfonts}
\usepackage{graphicx}

\begin{document}

\newcommand{\nn}{\nonumber}
\newcommand{\dg}{^\dagger}
\newcommand{\bra}[1]{\langle{#1}|}
\newcommand{\ket}[1]{|{#1}\rangle}
\newcommand{\braket}[2]{\langle{#1}|{#2}\rangle}

\title{Relation between classical communication capacity and entanglement
capability for two-qubit unitary operations}
\author{Dominic W.\ Berry and Barry C.\ Sanders}
\affiliation{Department of Physics and
        Centre for Advanced Computing -- Algorithms and Cryptography,   \\
        Macquarie University,
        Sydney, New South Wales 2109, Australia}
\date{\today}

\begin{abstract}
Two-qubit operations may be characterized by their capacities for communication, both
with and without free entanglement, and their capacity for creating entanglement.
We establish a set of inequalities that give an ordering to the capacities of
two-qubit unitary operations. Specifically, we show that the capacities for entanglement
creation and bidirectional communication without entanglement assistance are at least
as great as half the bidirectional communication capacity with entanglement
assistance. In addition, we show that the bidirectional communication that can be
performed using an ensemble may be increased via a two-qubit unitary operation by twice the
operation's capacity for entanglement.
\end{abstract}
\maketitle

\section{Introduction}
Quantum information processing relies on applications of single-qubit
unitary transformations and nonlocal two-qubit unitary transformations,
where the term ``nonlocal'' refers to joint operations on separate Hilbert
spaces. These nonlocal operations are powerful tools in quantum
information theory and can be characterized by the entanglement they can
generate between the two Hilbert spaces as well as by the communication
capacity they can deliver. These two seemingly disparate measures of gate
strengths are useful for quite different applications of two-qubit
nonlocal gates. However, we establish that these two capacities are closely
related.

Entanglement capability has been investigated in detail
\cite{cirac,zanardi,durvid,kradur,kraus,leifer,childs,kraus2,wang}.
The entanglement increase for finite unitaries
has been analyzed for initially unentangled states \cite{kraus} and
investigated by numerical methods for initially entangled states \cite{leifer}.
An important feature of the entanglement capability is that it quantifies the
``strength'' of the gate, which is particularly useful in considering the
resource requirements for simulation of specific gate operations or Hamiltonian
dynamics by other gates \cite{nielsen3}.

The communication capacity of nonlocal unitary transformations is as
important. The communication capacity is relevant to performing quantum
gates remotely using entanglement and classical communication
\cite{gotchu,eisert,collins,cirac,dur3}. An early example of the remarkable
relation between classical communication and entanglement was
seen in superdense coding \cite{super}.

It is known that there is a relationship between entanglement capability and
communication capacity for some simple gates. For example, the {\sc {\sc cnot}} gate can
create one ebit [the entanglement inherent in an ideal
Einstein-Podolsky-Rosen (EPR), or Bell, pair] of entanglement between
two initially unentangled systems; the {\sc cnot} gate can thus be used to
generate a Bell pair. In addition, the {\sc cnot} gate can be used to communicate
one bit of information in each direction simultaneously \cite{eisert,collins}.
The {\sc swap} gate may create two ebits of entanglement or perform two bits of
communication in one or both directions simultaneously
\cite{eisert,collins,super}. In both cases the communication is achieved while
consuming entanglement.

The relationship between communication capacity and entanglement
capability has been addressed by Bennett, Harrow, Leung, and Smolin (BHLS)
\cite{bennett}. They define asymptotic entanglement capabilities and
communication capacities for unitary operations and derive many relations for
these quantities.

Our goal here is to establish rigorous relations between entanglement
capability and communication capacity for two-qubit unitary operations. By
investigating the entanglement generated by a superposition of classical
messages, we recently showed that the entanglement capability of two-qubit
unitary operations is at least as great as half the capacity for error-free
entanglement-assisted communication \cite{PRL}. Here we provide an extended
version of those results, with rigorous derivations for nonzero error rate.

In Sec.\ \ref{comsch} we review the definitions of Ref.\ \cite{bennett}, with
minor modifications, and introduce the notation used throughout the paper. We
show that the communication rates may be achieved with an upper bound on the
dimension of the ancilla in Sec.\ \ref{limit}. We then apply this result to
prove an inequality between the entanglement-assisted communication capacity and
the capacities for creating and destroying entanglement (Sec.\ \ref{bipart}).

The results presented up to Sec.\ \ref{bipart} may be applied to all bipartite
unitary operations. We then restrict to two-qubit unitaries, and in
Secs.\ \ref{toocue} to \ref{ensemb} establish a sequence of inequalities between
the various capacities for two-qubit unitaries. We discuss a possible method
of proving equalities in Sec.\ \ref{ensemb}, and conclude in
Sec.\ \ref{conclude}.

\section{Capacity definitions}

\label{comsch}
The asymptotic definitions of capacities for entanglement creation and
communication were introduced by BHLS. In this section we summarize the
definitions of BHLS, and make some minor improvements. We consider two parties,
Alice and Bob, with the combined Hilbert space ${\cal H}_A\otimes{\cal H}_B$. We
call a unitary operator $U$ acting on ${\cal H}_A\otimes{\cal H}_B$ local if it
is of the form $V_A \otimes V_B$, where $V_A$ and $V_B$ act on ${\cal H}_A$ and
${\cal H}_B$, respectively. Operators that are not of this form we will term
nonlocal. (This is not meant to imply physical separation between the physical
systems represented by ${\cal H}_A$ and ${\cal H}_B$.) In general, the Hilbert
spaces ${\cal H}_A$ and ${\cal H}_B$ will be of the form
\begin{equation}
\label{ancilla}
{\cal H}_A = {\cal H}_{A_U} \otimes {\cal H}_{A_{\rm anc}}, ~~~~~
{\cal H}_B = {\cal H}_{B_U} \otimes {\cal H}_{B_{\rm anc}},
\end{equation}
such that $U$ acts only upon ${\cal H}_{A_U} \otimes {\cal H}_{B_U}$. The
Hilbert spaces ${\cal H}_{A_{\rm anc}}$ and ${\cal H}_{B_{\rm anc}}$ will be
called the ancillas.

The maximum increase in entanglement that may be produced by a single
implementation of the unitary transformation $U$ is denoted as $E_U$. That is,
we define
\begin{equation}
E_U \equiv \sup_{\ket{\Psi}\in{\cal H}_A\otimes{\cal H}_B}
\left[ E(U\ket{\Psi}) - E(\ket{\Psi})\right].
\end{equation}
The quantity $E(\cdots)$ is the entropy of entanglement
$E(\ket{\Psi}) = S[{\rm Tr}_A(\ket{\Psi}\bra{\Psi})]$, where
$S(\rho)=-{\rm Tr}(\rho \log\rho)$. Throughout we employ
logarithms to base 2, so the entanglement is expressed in units of ebits.
The ancillas may be of arbitrarily large, but finite, dimension. The
entanglement capability is maximized for pure states
\cite{kraus,bennett,leifer}, and we therefore restrict attention to pure states
in this paper. We also define the maximal decrease in entanglement
\begin{equation}
E_U^- \equiv \sup_{\ket{\Psi}\in{\cal H}_A\otimes{\cal H}_B}
\left[ E(\ket{\Psi}) - E(U\ket{\Psi})\right].
\end{equation}
BHLS show that the asymptotic entanglement capability per operation in the limit
of a large number of operations is equal to $E_U$. Similarly, $E_U^-$ may be
interpreted as the asymptotic entanglement capability of $U\dg$ \cite{bennett}.

To quantify the communication capacity of the unitary operation $U$, we adopt the
definitions of asymptotic capacities introduced by BHLS. The bidirectional
communication is quantified by the pair of numbers $(R_{\rightarrow},
R_{\leftarrow})$. The pair $(R_{\rightarrow},R_{\leftarrow})$ is said to be
achievable if, for any $\epsilon>0$, there exists $t$ such that it is possible
to communicate $tR_{\rightarrow}$ bits from Alice to Bob and $tR_{\leftarrow}$
bits from Bob to Alice with fidelity $1-\epsilon$ via $t$ applications of $U$
interspersed with local unitary operations.

To explain this communication scheme in more detail, Alice and Bob share the
Hilbert space ${\cal H}_A\otimes{\cal H}_B$, where ${\cal H}_A={\cal H}_{A_1}
\otimes{\cal H}_{A_2}$ and ${\cal H}_B={\cal H}_{B_1}\otimes{\cal H}_{B_2}$.
This division of the Hilbert space should not be confused with that in
Eq.\ \eqref{ancilla}; here we do not specify the component of the Hilbert space
upon which $U$ acts. Subsystems $A_1$ and $B_1$ are both of dimension $2^{\max
(n_a,n_b)}$. Subsystem $A_1$ is initially in one of $2^{n_a}$ orthogonal states
corresponding to Alice's $n_a$-bit message $x$, and $B_1$ is initially in a
state corresponding to Bob's $n_b$-bit message $y$. Subsystems $A_2$ and $B_2$
are the ancillas for Alice and Bob, respectively, and are initially in a shared
state $\ket{\psi}_{A_2B_2}$. The complete initial state may be expressed as
$\ket{x}_{A_1}\ket{y}_{B_1}\ket{\psi}_{A_2B_2}$.

The communication scheme consists of $t$ applications of $U$, interspersed with
the intermediate local unitary operations $V_A^{(k)}\otimes V_B^{(k)}$. The
final state is therefore
\begin{align}
\label{process}
\ket{\eta_{xy}}_{AB} &= (V_A^{(t)} \otimes V_B^{(t)}) U 
(V_A^{(t-1)} \otimes V_B^{(t-1)}) U \cdots 
\nn \\ & ~~~ \cdots U (V_A^{(0)} \otimes V_B^{(0)})
\ket{x}_{A_1}\ket{y}_{B_1} \ket{\psi}_{A_2B_2}.
\end{align}
For perfect communication Bob's message $y$ is transferred to $A_1$ and
Alice's message $x$ is transferred to $B_1$. For communication with fidelity at
least $1-\epsilon$, we have
\begin{equation}
F\left( \ket{y}_{A_1} \ket{x}_{B_1}, {\rm Tr}_{A_2B_2}
\ket{\eta_{xy}}_{AB}\bra{\eta_{xy}} \right) \ge 1-\epsilon,
\end{equation}
where $F(\ket{\psi},\rho)\equiv\bra{\psi}\rho\ket{\psi}$. This condition means
that the classical error rate is no greater than $\epsilon$.

As in Ref.\ \cite{bennett}, we say that the pair $(R_{\rightarrow},
R_{\leftarrow})$ is achievable if, for all $\epsilon>0$, there exists a $t$ such
that it is possible to perform communication of fidelity at least $1-\epsilon$
in this way with $n_a\ge tR_\rightarrow$ and $n_b\ge tR_\leftarrow$.

For communication that is not assisted by entanglement, we take the initial
ancilla state $\ket{\psi}_{A_2B_2}$ to be an unentangled state of arbitrary
finite dimension. For entanglement-assisted communication, we allow an ancilla
state that is a tensor product of an unentangled state and a maximally entangled
state, both of which may be of arbitrary finite dimension.

For the entanglement-assisted case, it is necessary to allow an unentangled
component to the ancilla because the communication scheme is composed
entirely of unitary operations. Some of the processes that communication schemes
use require an unentangled subsystem, and it is not possible to obtain an
unentangled subsystem using local unitary operations.

For example, consider the measurement of some state $\ket{\psi}$ via a
$d$-element positive operator-valued measure $\{\Pi_i\}$. In order to treat
the measurement as a unitary process, we add a register of dimension $d$,
which is initialized in an unentangled state $\ket{0}$. The measurement may
then be represented by a unitary operation $U_{\rm meas}$ such that
$U_{\rm meas} \ket{0}\ket{\psi} = \sum_i \ket{i}\sqrt{\Pi_i}\ket{\psi}$.
The register stores the result of the
measurement, which is only possible because it is initially unentangled.

Thus it is necessary to have an unentangled subsystem in order to perform the
unitary equivalent of local measurements. In particular, because the
communication scheme considered by Ref.\ \cite{bennett} uses local measurements,
it requires an unentangled resource. It is therefore necessary to allow an
unentangled component to the ancilla for our definition to be consistent with
Ref.\ \cite{bennett}.

We denote the set of achievable rate pairs without entanglement assistance by
$S_U$, and the set of achievable rate pairs with entanglement assistance by
$S_U^E$. We will also use a superscript $X$ to indicate both cases. The explicit
expressions for $S_U$ and $S_U^E$ are given in Appendix \ref{def}. The
communication rates for the bidirectional and unidirectional cases are then
defined by
\begin{align}
C_+^X &= \sup_{(R_{\rightarrow},R_{\leftarrow})\in S_U^X}
(R_{\rightarrow}+R_{\leftarrow}), \\
C_{\rightarrow}^X & = \sup_{(R_{\rightarrow},R_{\leftarrow})\in S_U^X}
R_{\rightarrow}, ~~~ 
C_{\leftarrow}^X = \sup_{(R_{\rightarrow},R_{\leftarrow})\in S_U^X}
R_{\leftarrow}.
\end{align}
Here we have used the $X$ superscript to indicate that the same definitions hold
with entanglement assistance and without entanglement assistance.

\section{Limiting the ancilla}
\label{limit}
In the above definitions, the size of the ancilla is always finite, but has no
upper limit, and can grow arbitrarily quickly with $t$. This unbounded nature of
the dimension of the ancilla causes difficulties in deriving results for the
communication capacities. To avoid these problems, in this section we show how
to put an upper bound on the dimension of the ancilla; that is, we show that for
all achievable rate pairs (that are not on the boundary of the region of
achievable rate pairs), there exists a constant $K$ such that this rate pair may
be achieved with an ancilla space of dimension no larger than $2^{Kt}$.

In Ref.\ \cite{bennett} similar results were derived for communication without
prior entanglement and for one-way communication with entanglement.
Here we give a general derivation that may be applied to these cases, and also
to bidirectional communication with entanglement.

The set of achievable rate pairs $(R_{\rightarrow}, R_{\leftarrow})$, $S_U^X$,
forms a two-dimensional region with an upper boundary \cite{bennett}. The interior
of $S_U^X$ is an open set; therefore, for any rate pair
$(R_{\rightarrow},R_{\leftarrow})$ in the interior of $S_U^X$, there must be an
achievable pair $(R'_{\rightarrow},R'_{\leftarrow})$
closer to the upper boundary, so $R'_{\rightarrow}>R_{\rightarrow}$ and
$R'_{\leftarrow}>R_{\leftarrow}$. These rate pairs may
be for the entanglement assisted or unassisted case; that is, our
derivation applies to both cases.

It is possible to achieve the rate pair $(R'_{\rightarrow},R'_{\leftarrow})$
with fidelity $1-\epsilon$, where
\begin{equation}
\label{epsilon}
\epsilon \le \min\left\{ \left( \frac{\Delta R} {32R_{\rm max}}
\right)^2, \left( \frac{\Delta R}{16} \right)^4,
\left(\frac 1{2e}\right)^2 \right\},
\end{equation}
with
\begin{equation}
\Delta R = \min \left\{ R'_{\rightarrow} - R_{\rightarrow} ,
R'_{\leftarrow} - R_{\leftarrow} \right\},
\end{equation}
and $R_{\rm max} = \max \{ R'_{\rightarrow}, R'_{\leftarrow},1 \}$.
The reason for putting this restriction on $\epsilon$ will become clear below.
This communication may be achieved with an ancilla of some finite dimension,
$d$, and $\tau$ applications of $U$. Without loss of generality we restrict the
length of the messages communicated, $n_a$ and $n_b$, to be no longer than
$2\tau R_{\rm max}$ bits. Any excess bits are unnecessary for achieving the
communication rate, and will be included as part of the ancilla. We may reduce
the error rate by coding over $M$ blocks of $\tau$ operations, for a total
number of operations equal to $t=M\tau$.

For Bob's message $y$, Bob may take Alice's final reduced density matrix in
$A_1$ to be
\begin{equation}
\rho_{A_1}^{(y)} = 2^{-n_a} \sum_{x=0}^{2^{n_a}-1} {\rm Tr}_{A_2B}
(\ket{\eta_{xy}}_{AB}\bra{\eta_{xy}}).
\end{equation}
Here we have averaged over $x$, as Bob does not have this information. Using the
concavity of the fidelity and the fact that the fidelity is nondecreasing under
partial trace \cite{nielsen}, the fidelity of $\rho_{A_1}^{(y)}$ with
$\ket{y}_{A_1}$ is at least $1-\epsilon$. Similarly, Alice may take Bob's density
matrix to be
\begin{equation}
\rho_{B_1}^{(x)} = 2^{-n_b} \sum_{y=0}^{2^{n_b}-1} {\rm Tr}_{AB_2}
(\ket{\eta_{xy}}_{AB}\bra{\eta_{xy}}).
\end{equation}
This will have fidelity at least $1-\epsilon$ with $\ket{x}_{B_1}$.

When density matrix $\rho_i$ is given with probability $p_i$, we denote this
by the ensemble ${\sf E}=\{p_i,\rho_i\}$. From Refs.\ \cite{holevo,shuwes}, the
communication that may be performed with such an ensemble is given by the
Holevo information
\begin{equation}
\label{holevo}
\chi({\sf E})=S\left( \sum_i p_i \rho_i \right) - \sum_i p_i S(\rho_i).
\end{equation}
This communication may be obtained
asymptotically by using error correcting codes over multiple states.

For the communication scheme above, the ensemble possessed by Alice is
${\sf E}_A=\{ 2^{-n_b},\rho_{A_1}^{(y)} \}$. The elements of this ensemble have
fidelity at least $1-\epsilon$ with the ensemble ${\sf E}_A^0=\{ 2^{-n_b},
\ket{y}_{A_1}\bra{y} \}$. Using the concavity of the fidelity, the average
density matrix $\bar \rho_{A_1} = 2^{-n_b}\sum_y \rho_{A_1}^{(y)}$ has
fidelity at least $1-\epsilon$ with the average density matrix for
${\sf E}_A^0$ given by $\bar \sigma_{A_1}=2^{-n_b}\sum_y \ket{y}_{A_1}\bra{y}$.

It is easily shown that the Holevo information for ${\sf E}_A^0$ is $n_b$.
To estimate the Holevo information for ${\sf E}_A$ we may use Fannes'
inequality \cite{fannes}
\begin{equation}
\label{fan}
|S(\rho)-S(\sigma)|\le T(\rho,\sigma)\log D+\eta(T(\rho,\sigma)),
\end{equation}
where $D$ is the dimension of the Hilbert space and $\eta(x)\equiv-x\log x$.
$T(\rho,\sigma)\equiv{\rm Tr}|\rho-\sigma|$ is the trace distance, and we
require that $T(\rho,\sigma)\le 1/e$.

For $\rho=\rho_{A_1}^{(y)}$ and $\sigma=\ket{y}_{A_1}\bra{y}$, or
$\rho=\bar \rho_{A_1}$ and $\sigma=\bar \sigma_{A_1}$, the fidelity
is at least $1-\epsilon$. Using the inequality between trace distance and
fidelity given in Ref.\ \cite{nielsen}, this implies that $T(\rho,\sigma) \le
2\sqrt\epsilon$. The fact that we have taken $\epsilon\le (1/2e)^2$ implies
that $T(\rho,\sigma)\le 1/e$ is satisfied.

Using these results, and the fact that $\eta(x)\le \sqrt{2x}$, the Holevo
information for ${\sf E}_A$ must satisfy
\begin{equation}
\chi({\sf E}_A) \ge n_b-4n\sqrt\epsilon-4\epsilon^{1/4},
\end{equation}
where $n=\max\{n_a,n_b\}$. Using the inequalities $n\le 2\tau R_{\rm max}$,
$n_b\ge\tau R'_\leftarrow$, $\tau\ge 1$ and the restrictions put on
$\epsilon$ in Eq.\ \eqref{epsilon}, we get
\begin{equation}
\chi({\sf E}_A) \ge \tau (R'_\leftarrow+R_\leftarrow)/2 .
\end{equation}
From Ref.\ \cite{holevo}, by coding over a number of ensembles $M$ no less than
some minimum $M_A$, it is possible to communicate information
per ensemble from Bob to Alice arbitrarily close to $\chi({\sf E}_A)$ with
average fidelity arbitrarily close to 1. For example, we may achieve
communication $M\tau(R'_\leftarrow+3R_\leftarrow)/4$ with average fidelity
$1-\tfrac 12\epsilon_0(R'_\leftarrow-R_\leftarrow)/(R'_\leftarrow+3
R_\leftarrow)$. The fact that the result of Ref.\ \cite{holevo} is given in
terms of the average fidelity, rather than the minimum fidelity, is unimportant
because the codewords that result in low fidelity may be omitted. For this
example, no more than a proportion $(R'_\leftarrow-R_\leftarrow)/
(R'_\leftarrow+3R_\leftarrow)$ of the codewords may give fidelity less
than $1-\epsilon_0/2$. Omitting these then gives communication of $M\tau
R_\leftarrow$ with fidelity of at least $1-\epsilon_0/2$.

The same considerations hold for communication from Alice to Bob. For Bob's
ensemble we find
\begin{equation}
\chi({\sf E}_B) \ge \tau (R'_\rightarrow+R_\rightarrow)/2 .
\end{equation}
Similarly, by coding over a large number $M\ge M_B$ of ensembles, it is possible
to achieve communication from Alice to Bob of $M\tau R_\rightarrow$ with
fidelity of at least $1-\epsilon_0/2$. If we take the number of ensembles $M$ to
be larger than both $M_A$ and $M_B$, then it is possible to perform the coding
for communication in each direction on the same block of $M$ states. In this
way, communication $M\tau R_\leftarrow$ from Bob to Alice and
$M\tau R_\rightarrow$ from Alice to Bob may be performed, with the total
fidelity at least $1-\epsilon_0$.

Thus we see that, by coding over $M$ blocks of $\tau$ applications of $U$, we
may perform communication per operation of $R_\leftarrow$ from Bob to Alice
and $R_\rightarrow$ from Alice to Bob with arbitrarily high fidelity. $M$ copies
of the individual ancillas $\ket{\psi}_{A_2B_2}$ are required, each of dimension
$d$, for a total dimension of $d^M$. In addition, we should take account of the
$M$ copies of subsystems $A_1$ and $B_1$. These subsystems contain the
codewords, and the decoded messages will be contained in separate subsystems.
We therefore find that the dimension of the ancilla required for the total
communication scheme is no more than $2^{Kt}$, where $t=M\tau$ is the total
number of operations and $K=(2n+\log d)/\tau$.

Thus we see that any achievable rate pair $(R_{\rightarrow},R_{\leftarrow})$
in the interior of $S_U^X$ may be achieved with a bound $2^{Kt}$ on the
dimension of the ancilla. The restriction that the rate pair must be in the
interior of $S_U^X$ does not affect the results for the communication capacities,
because these capacities are defined via a supremum, and we may take the rate
pair to be arbitrarily close to the upper boundary.

Note also that, because we have taken $n_a$ and $n_b$ for the individual
communication schemes (with $\tau$ applications of $U$) to be no larger than
$2\tau R_{\rm max}$, the total communication in either direction may
be bounded by $tK_n$, where $K_n=2R_{\rm max}$. This bound is
relatively trivial, but will be used later. Another application of these results
is that $t$ may be taken to approach infinity in the limit $\epsilon\to 0$,
which is not explicitly given in the definition of the rates.

\section{Communication with entanglement}
\label{entang}
The fact that it is possible to limit the dimension of the ancilla in this way
considerably simplifies some of the analysis in Ref.\ \cite{bennett}, and
allows additional relations to be derived. It is possible to derive
inequalities between the classical communication capacity and the entanglement
capability using a quantum superposition of classical messages. However, there
is the complication that the entanglement only converges if there is an
upper limit placed on the dimension.

In Ref.\ \cite{bennett} the inequalities $E_U \ge C_+$ and
$E_U+E_U^-\ge C^E_\rightarrow$ were proven. The ancillas were bounded using
techniques that are specific to these inequalities and cannot be generalized.
The bound that we have placed on the dimension of the ancilla in
Sec.\ \ref{limit} is sufficiently general to be applied to show these
inequalities, and show further inequalities. In Sec.\ \ref{bipart} below, we
show the additional inequality $E_U+E_U^-\ge C_+^E$. This inequality is for
general bipartite unitary operations, and is not restricted to two-qubit unitaries.
In Sec.\ \ref{toocue} we show a simplified inequality that is restricted to
two-qubit unitary operations.

\subsection{Arbitrary bipartite unitary operations}
\label{bipart}
Let $(R_\rightarrow,R_\leftarrow)$ be a pair of rates in the interior of $S_U^E$.
From Sec.\ \ref{comsch}, there exists a protocol that transmits
$n_a\ge tR_\rightarrow$ bits from Alice to Bob and $n_b\ge t R_\leftarrow$ bits
from Bob to Alice, via $t$ uses of $U$ and with fidelity $1-\epsilon$. From
Sec.\ \ref{limit}, there exists a constant $K$ such that $\epsilon$ may be made
arbitrarily small while the dimension of the ancilla is no greater than
$2^{Kt}$. The final state $\ket{\eta_{xy}}_{AB}$ satisfies
\begin{equation}
F\left( \ket{y}_{A_1}\ket{x}_{B_1},{\rm Tr}_{A_2B_2}\ket{\eta_{xy}}_{AB}
\bra{\eta_{xy}}\right)=1-\epsilon_{xy},
\end{equation}
with $\epsilon_{xy}\le\epsilon$ for all messages $x$ and $y$. Using Uhlmann's
theorem, the state $\ket{\eta_{xy}}_{AB}$ may be expressed as \cite{uhlmann}
\begin{equation}
\label{uhl}
\ket{\eta_{xy}}_{AB} = \sqrt{1-\epsilon_{xy}}\ket{y}_{A_1}\ket{x}_{B_1}
\ket{c_{xy}}_{A_2B_2}+\sqrt{\epsilon_{xy}}\ket{e_{xy}}_{AB},
\end{equation}
where ${\rm Tr}_{A_2B_2}\ket{e_{xy}}_{AB}\bra{e_{xy}}$ has support orthogonal
to $\ket{y}_{A_1}\ket{x}_{B_1}$. The change in the entanglement is
\begin{equation}
\Delta E_{xy} = E(\ket{\eta_{xy}}_{AB}) - E(\ket{\psi}_{A_2B_2} ).
\end{equation}

Now we add additional subsystems $A_3$ and $B_3$ that contain copies of Alice's
message $x$ and Bob's message $y$. In addition, rather than selecting
specific classical messages $x$ and $y$, we have a coherent superposition of
all possible messages. Then the input state is
\begin{equation}
2^{-(n_a+n_b)/2}\sum_{xy} \ket{x}_{A_1}\ket{x}_{A_3}\ket{y}_{B_1}\ket{y}_{B_3}
\ket{\psi}_{A_2B_2},
\end{equation}
and the output state is
\begin{equation}
\label{output}
\ket{\eta_\epsilon} = 2^{-(n_a+n_b)/2}\sum_{xy} \ket{x}_{A_3} \ket{y}_{B_3}
\ket{\eta_{xy}}_{A_{12}B_{12}}.
\end{equation}
Here we use the notation of multiple subscripts to indicate the combined
Hilbert space, for example, $A_{12}$ indicates ${\cal H}_{A_1}\otimes
{\cal H}_{A_2}$. We use $A$ and $B$ to indicate the entire Hilbert spaces of
Alice and Bob, respectively; so, for example, $A$ is equivalent to $A_{123}$.

In order to put bounds on the entanglement that may be created, we compare
Eq.\ \eqref{output} with the state
\begin{equation}
\ket{\eta}=2^{-(n_a+n_b)/2}\sum_{xy} \ket{y}_{A_1}\ket{x}_{A_3}\ket{x}_{B_1}
\ket{y}_{B_3}\ket{c_{xy}}_{A_2B_2}.
\end{equation}
We find that
\begin{align}
\braket{\eta}{\eta_\epsilon} &= 2^{-(n_a+n_b)}\sum_{xy}(_{A_1}\bra{y}_{B_1}
\bra{x}_{A_2B_2}\bra{c_{xy}})\ket{\eta_{xy}}_{A_{12}B_{12}} \nn \\
& \ge 2^{-(n_a+n_b)}\sum_{xy} \sqrt{1-\epsilon} = \sqrt{1-\epsilon}, 
\end{align}
giving $|\braket{\eta}{\eta_\epsilon}|^2\ge 1-\epsilon$.

From the continuity of the entropy of entanglement \cite{contin}, or Fannes'
inequality \cite{fannes}, we find that the difference between the entanglement
of the two states $\ket{\eta}$ and $\ket{\eta_\epsilon}$ is bounded by
\begin{align}
|E(\ket{\eta})-E(\ket{\eta_\epsilon})| \le T(\ket{\eta},\ket{\eta_\epsilon})
(2n+Kt/2)+Q,
\end{align}
where $n=\max(n_a,n_b)$, $Q=\log(e)/e$ and $T(\ket{a},\ket{b})={\rm Tr}
\big|\ket{a}\bra{a}-\ket{b}\bra{b}\big|$ is the trace distance. Here we have
used the fact that the dimension of the ancilla is bounded by $2^{Kt}$. Using
the inequality between trace distance and fidelity given in
Ref.\ \cite{nielsen}, we find that
\begin{align}
|E(\ket{\eta})-E(\ket{\eta_\epsilon})| & \le \sqrt{1-|\braket{\eta}
{\eta_\epsilon}|^2}(4n +Kt) + Q \nn \\
& \le \sqrt{\epsilon}(4n +Kt)+Q .
\end{align}

Therefore the increase in the entanglement for $t$ applications of $U$ is
\begin{align}
\label{ineq}
\Delta E & = E(\ket{\eta_\epsilon})-E(\ket{\psi}_{A_2B_2}) \nn \\
& \ge E(\ket{\eta})-\sqrt{\epsilon}(4n+Kt)-Q-E(\ket{\psi}_{A_2B_2}).
\end{align}
The entanglement of $\ket{\eta}$ may be found, using Eq.\ (11.58) of
Ref.\ \cite{nielsen}, to be
\begin{equation}
E(\ket{\eta}) = n_a+n_b +2^{-(n_a+n_b)}\sum_{xy}E(\ket{c_{xy}}_{A_2B_2}).
\end{equation}
Substituting this into Eq.\ \eqref{ineq} gives
\begin{align}
\label{almost}
& \Delta E \ge n_a+n_b - \sqrt{\epsilon}(4n+Kt)-Q \nn \\
& ~~~ +2^{-(n_a+n_b)}\sum_{xy}\left[ E(\ket{c_{xy}}_{A_2B_2})
- E(\ket{\psi}_{A_2B_2}) \right].
\end{align}

Again applying the continuity of the entropy of entanglement \cite{contin},
\begin{align}
& \left| E(\ket{\eta_{xy}}_{A_{12}B_{12}}) - E(\ket{y}_{A_1}\ket{x}_{B_1}
\ket{c_{xy}}_{A_2B_2})\right| \nn \\
& \le T(\ket{\eta_{xy}}_{A_{12}B_{12}},\ket{y}_{A_1}\ket{x}_{B_1}
\ket{c_{xy}}_{A_2B_2})(n +Kt/2) + Q \nn \\
& \le \sqrt{\epsilon}(2n +Kt) + Q.
\end{align}
Using this, and the fact that $\Delta E_{xy}\ge -tE_U^-$, we find
\begin{align}
& E(\ket{c_{xy}}_{A_2B_2}) - E(\ket{\psi}_{A_2B_2}) \nn \\
& \ge E(\ket{\eta_{xy}}_{A_{12}B_{12}}) - \sqrt{\epsilon}(2n+Kt)
- Q -E(\ket{\psi}_{A_2B_2} ) \nn \\
& \ge -tE_U^- - \sqrt{\epsilon}(2n+Kt) - Q .
\end{align}
Substituting this into Eq.\ \eqref{almost}, and noting that
$\Delta E \le tE_U$, gives
\begin{align}
tE_U & \ge n_a+n_b -tE_U^- - 2\sqrt{\epsilon}[3n+Kt]-2Q.
\end{align}
Using the inequalities $n \le tK_n$ (from Sec.\ \ref{limit}),
$n_a\ge tR_\rightarrow$, $n_b\ge tR_\leftarrow$, and dividing on both sides
by $t$, gives
\begin{align}
\label{above}
E_U + E_U^- & \ge R_\rightarrow+R_\leftarrow - 2\sqrt{\epsilon}[3K_n+K]-2Q/t.
\end{align}

In the limit $\epsilon\to 0$, as $K$ and $K_n$ are fixed, the third term on the
right-hand side goes to zero. The fourth term goes to zero in this limit,
because $t$ may be taken to approach infinity (as mentioned in the preceding
section). We therefore obtain $E_U+E_U^-\ge R_\rightarrow+R_\leftarrow$, for
any achievable pair of rates $(R_\rightarrow,R_\leftarrow)$ that is in the
interior of $S_U^E$. This implies that
\begin{equation}
E_U+E_U^- \ge C_+^E.
\end{equation}

Note that it is the upper limit $K$ on the number of auxiliary qubits that may
be used per operation that allows us to take the limit $\epsilon\to 0$. When
the dimension of the system is limited in this way, the entanglement has good
continuity properties. This continuity means that we obtain the same result
in the limit $\epsilon\to 0$ as we would if the communication were exact
($\epsilon=0$).

\subsection{Two-qubit unitary operations}
\label{toocue}
For the case of two-qubit unitaries, we may simplify this result. In this case
the maximum increase in the entanglement $E_U$ is equal to the maximum decrease
in the entanglement $E_U^-$, which may be shown in the following way. Any
two-qubit interaction $U$ is equivalent, up to local unitary operations, to an
operation of the form \cite{durvid,kraus,makhlin,hammer}
\begin{equation}
\label{simpleU}
U_d(\alpha_1,\alpha_2,\alpha_3) = e^{ -i \left( \alpha_1 \sigma_1 \otimes
\sigma_1 + \alpha_2 \sigma_2 \otimes \sigma_2 + \alpha_3 \sigma_3 \otimes
\sigma_3 \right) },
\end{equation}
where $\sigma_1$, $\sigma_2$, and $\sigma_3$ are the Pauli sigma matrices.

As entanglement capability and classical communication capacity are independent
of local unitary operations, we may restrict to operations of this form. It is
then simple to show that $U_d^*=U_d\dg=U_d^{-1}$. As discussed in
Ref.\ \cite{kraus}, for any measure of entanglement,
$E(\ket{\Psi})=E(\ket{\Psi^*})$. This means
that, if the operation $U_d$ acting on the state $\ket{\Psi}$ generates the
maximum increase in entanglement, then this operation performed on the state
$U_d^*\ket{\Psi^*}$ decreases the entanglement by $E_U$. Therefore, the
operation may decrease the entanglement at least as much as it may increase it.
Similarly it is simple to show the converse, and therefore $E_U=E_U^-$. Thus we
find that, for two-qubit unitary operations, we have the inequality
\begin{equation}
\label{myineq}
2E_U \ge C_+^E.
\end{equation}

\section{Communication without entanglement}
\label{unass}
For the case of two-qubit unitaries we may obtain inequalities between
the communication that may be obtained with and without entanglement. To see
this, again consider any pair of rates $(R_\rightarrow,R_\leftarrow)$ in
the interior of $S_U^E$. We may select a second pair
$(R'_\rightarrow,R'_\leftarrow)$ that is closer to but not on the upper
boundary, so that $R'_{\rightarrow}>R_{\rightarrow}$ and
$R'_{\leftarrow}>R_{\leftarrow}$.

Then there exists a communication scheme that communicates
$n_a\ge \tau R'_\rightarrow$ bits from Alice to Bob and
$n_b\ge \tau R'_\leftarrow$ bits from Bob to Alice via $\tau$ uses of $U$ and
with fidelity $1-\epsilon$. The input state is $\ket{x}_{A_1}\ket{y}_{B_1}
\ket{\psi}_{A_2B_2}$, and the output state $\ket{\eta_{xy}}_{AB}$ is obtained
via a process of the form of Eq.\ \eqref{process}.

The process described by Eq.\ \eqref{process} consumes entanglement resources to
perform this communication. We consider a strategy for recovering the initial
entanglement resource state $\ket{\psi}_{A_2B_2}$ by subsequently performing the
communication scheme \eqref{process} \emph{in reverse}. (This approach is
analogous to that applied in Ref.\ \cite{cleve}.) To achieve this, Alice and
Bob must again retain copies of their inputs in auxiliary subsystems $A_3$ and
$B_3$. In addition, to ensure that at the end of this process Alice and Bob
retain the communicated information, Bob must create a copy of the output $x$
and Alice must create a copy of $y$. We therefore add auxiliary subsystems
$A_4$ and $B_4$ that are initially in the state $\ket{0}$. After performing the
communication we copy the value $y$ from $A_1$ to $A_4$ and copy the value $x$
from $B_1$ to $B_4$.

From Eq.\ \eqref{uhl}, we may express the state after the communication
process, but before the outputs are copied into $A_4$ and $B_4$, as
\begin{align}
&\left\{ \sqrt{1-\epsilon_{xy}}\ket{y}_{A_1}\ket{x}_{B_1}\ket{c_{xy}}_{A_2B_2}+
\sqrt{\epsilon_{xy}}\ket{e_{xy}}_{A_{12}B_{12}} \right\} \nn \\ & ~~~~~~
\otimes \ket{x}_{A_3}\ket{y}_{B_3}\ket{0}_{A_4}\ket{0}_{B_4},
\end{align}
where $\ket{e_{xy}}_{A_{12}B_{12}}$ is some normalized error state. Copying the
results of the communication to the auxiliary subsystems $A_4$ and $B_4$ then
yields
\begin{align}
\label{finst}
\sqrt{1-\epsilon_{xy}}\ket{y}_{A_1}\ket{x}_{B_1}\ket{c_{xy}}_{A_2B_2}
\ket{y}_{A_4}\ket{x}_{B_4} \ket{x}_{A_3}\ket{y}_{B_3} \nn \\
+\sqrt{\epsilon_{xy}}\ket{e^{(1)}_{xy}}_{A_{124}B_{124}}\ket{x}_{A_3}
\ket{y}_{B_3}.
\end{align}
Here we are using superscripts on the error state to indicate that it has also
been changed under this operation.

In order to reverse the communication scheme, we wish to apply the inverse of
the sequence of operations that was used to perform the communication:
\begin{align}
\label{normseq}
(V_{A_{12}}^{(0)}\otimes V_{B_{12}}^{(0)})\dg U\dg \cdots U\dg
(V_{A_{12}}^{(\tau)} \otimes V_{B_{12}}^{(\tau)})\dg.
\end{align}
It is possible to apply the inverses of the local operations $V_{A_{12}}^{(k)}$
and $V_{B_{12}}^{(k)}$, because we assume that it is possible to perform
arbitrary local unitary operations. Applying the inverse of operation $U$ is
more difficult, however. When the operation $U$ is in the simple form
\eqref{simpleU}, the inverse operation $U\dg$ is equal to $U^*$. If we take the
complex conjugate of the state, then the action of $U$ on this complex
conjugate state will be the same as that of $U^*$ on the original state. To
reverse the communication scheme on the complex conjugate state, we simply
apply the complex conjugate of the sequence of operations in
Eq.\ \eqref{normseq}:
\begin{align}
\label{seq}
(V_{A_{12}}^{(0)}\otimes V_{B_{12}}^{(0)})^T U \cdots U
(V_{A_{12}}^{(\tau)}\otimes V_{B_{12}}^{(\tau)})^T.
\end{align}

Here we can not take the exact complex conjugate of the state, but we can take
the complex conjugate of the first term in Eq.\ \eqref{finst}. To obtain
the complex conjugate of this term, note that we may express
$\ket{c_{xy}}_{A_2B_2}$, via a Schmidt decomposition, in the form
\begin{equation}
\ket{c_{xy}}_{A_2B_2} = \sum_i \sqrt{\lambda_{xy}^i}\ket{\varphi_{xy}^i}_{A_2}
\ket{\chi_{xy}^i}_{B_2}.
\end{equation}
The first term in Eq.\ \eqref{finst} is therefore proportional to the state
\begin{equation}
\ket{y}_{A_1}\ket{x}_{A_3}\ket{y}_{A_4}\ket{x}_{B_1}\ket{y}_{B_3}\ket{x}_{B_4}
\! \sum_i \! \sqrt{\lambda_{xy}^i}\ket{\varphi_{xy}^i}_{A_2}
\ket{\chi_{xy}^i}_{B_2}.
\end{equation}
We may take the complex conjugate of this state using the local operations
\begin{align}
\label{conjops}
&\sum_{xy} \ket{y}_{A_1}\bra{y} \otimes \ket{x}_{A_3}\bra{x} \otimes
\sum_i \ket{\varphi_{xy}^{i*}}_{A_2}\bra{\varphi_{xy}^i}, \\
&\sum_{xy} \ket{x}_{B_1}\bra{x} \otimes \ket{y}_{B_3}\bra{y} \otimes
\sum_i \ket{\chi_{xy}^{i*}}_{B_2}\bra{\chi_{xy}^i}.
\end{align}
These are conditional operations with $A_1$ and $A_3$ (for Alice) and
$B_1$ and $B_3$ (for Bob) as the controls. Note that it is retaining the value
of $x$ in $A_3$ and the value of $y$ in $B_3$ that makes these operations
possible.

Applying these operations gives the state
\begin{align}
\label{conjstate}
\sqrt{1-\epsilon_{xy}}\ket{y}_{A_1}\ket{x}_{B_1}\ket{c^*_{xy}}_{A_2B_2}
\ket{y}_{A_4}\ket{x}_{B_4}\ket{x}_{A_3}\ket{y}_{B_3} \nn \\
+\sqrt{\epsilon_{xy}}\ket{e^{(2)}_{xy}}_{A_{124}B_{124}} \ket{x}_{A_3}
\ket{y}_{B_3}.
\end{align}
Now we have a state that is close to the complex conjugate of the output state
of the communication process. It is therefore possible to approximately
reverse the communication process by performing the sequence of operations
given in Eq.\ \eqref{seq} to state \eqref{conjstate}. In this way, via
$2\tau$ implementations of $U$, we obtain the output state
\begin{align}
& \left\{ \left[ \ket{x}_{A_1}\ket{y}_{B_1} \ket{\psi^*}_{A_2B_2} -
\sqrt{\epsilon_{xy}}\ket{e^{(3)}_{xy}}_{A_{12}B_{12}} \right]
\ket{y}_{A_4}\ket{x}_{B_4} \right. \nn \\ & ~~~~ \left. +\sqrt{\epsilon_{xy}}
\ket{e^{(4)}_{xy}}_{A_{124}B_{124}} \right\} \ket{x}_{A_3}\ket{y}_{B_3}.
\end{align}
Then we perform local transformations that take the complex conjugate of
$\ket{\psi^*}_{A_2B_2}$, giving the final state
\begin{align}
\label{error2}
\ket{\zeta_1}&=\ket{x}_{A_1}\ket{x}_{A_3}\ket{y}_{B_1}\ket{y}_{B_3}
\ket{\psi}_{A_2B_2}\ket{y}_{A_4}\ket{x}_{B_4} \nn \\ &~~~ -\sqrt{\epsilon_{xy}}
\ket{e^{(5)}_{xy}}+\sqrt{\epsilon_{xy}}\ket{e^{(6)}_{xy}}.
\end{align}
Here we have omitted the subscript $AB$ on states in the entire Hilbert space
${\cal H}_A\otimes{\cal H}_B$ for brevity.

We now consider a sequence of $M$ processes of this type, communicating $M$
messages $x_1$ to $x_M$ from Alice to Bob and $M$ messages $y_1$ to $y_M$ from
Bob to Alice. If we use the output state from the first communication process
\eqref{error2}, and perform the same communication process to communicate the
second two messages $x_2$ and $y_2$, then the state we will obtain is
\begin{align}
\ket{\zeta_2} &=\ket{x_1x_2}_{A_1}\ket{x_1x_2}_{A_3}\ket{y_1y_2}_{B_1}
\ket{y_1y_2}_{B_3}\ket{\psi}_{A_2B_2}\ket{y_1y_2}_{A_4} \nn \\ & ~~~ \otimes
\ket{x_1x_2}_{B_4}-\sqrt{\epsilon_{x_1 y_1}}\ket{e^{(7)}_{x_1 y_1}}
+\sqrt{\epsilon_{x_1 y_1}}\ket{e^{(8)}_{x_1 y_1}} \nn \\ & ~~~
-\sqrt{\epsilon_{x_2 y_2}}\ket{e^{(5)}_{x_2 y_2}}
+\sqrt{\epsilon_{x_2 y_2}}\ket{e^{(6)}_{x_2 y_2}}.
\end{align}

Doing this $M$ times, each step generates an additional two error terms of the
form given in Eq.\ \eqref{error2}, resulting in the final state
\begin{align}
\ket{\zeta_M}&=\ket{X}_{A_1}\ket{X}_{A_3}\ket{Y}_{B_1}\ket{Y}_{B_3}
\ket{\psi}_{A_2B_2}\ket{Y}_{A_4}\ket{X}_{B_4} \nn \\ &~~~ +\sum_{i=1}^M
\sqrt{\epsilon_{x_i y_i}}\left(-\ket{e^{(5,i)}_{x_i y_i}} +
\ket{e^{(6,i)}_{x_i y_i}} \right) .
\end{align}
Here we have used $X$ to represent the sequence of messages
$(x_1,x_2,\ldots,x_M)$ and $Y$ to represent the sequence of messages
$(y_1,y_2,\ldots,y_M)$.
In order to perform perfect communication we would require the state
\begin{equation}
\ket{\zeta}=\ket{X}_{A_1}\ket{X}_{A_3}\ket{Y}_{B_1}\ket{Y}_{B_3}
\ket{\psi}_{A_2B_2}\ket{Y}_{A_4}\ket{X}_{B_4}.
\end{equation}
It is easy to see that the fidelity of the state $\ket{\zeta_M}$ with respect to
the state for perfect communication satisfies
\begin{equation}
|\braket{\zeta}{\zeta_M}|^2 \ge 1-4M\sqrt{\epsilon}.
\end{equation}

In order to communicate a large amount of information without prior
entanglement, we construct an approximation of the initial resource state
$\ket{\psi}_{A_2B_2}$ using a limited number of applications of $U$. We again
use the fact that we are limiting the dimension of the auxiliary state that
may be used. The dimension of $\ket{\psi}_{A_2B_2}$ may not be larger than
$2^{K\tau}$, which means that the entanglement of this state cannot be larger
than $K\tau/2$ ebits. It is easily seen that this state may be created with
$O(K\tau)$ operations.

In particular, let $\ket{\psi_0}$ be a state with entanglement $E_0>0$ that can
be created via a single application of $U$ from an initially unentangled
state. From the theory of entanglement concentration \cite{BBPS}, we may create
a maximally entangled state with entanglement up to $K\tau/2$ and fidelity
at least $1-\epsilon_{\psi}$ using $K\tau/2E_0+C(\epsilon_{\psi},\psi_0)
\sqrt{K\tau}$ copies of the state $\ket{\psi_0}$. Here $C(\epsilon_{\psi},
\psi_0)$ is a constant that depends on the fidelity required and the initial
state. Thus the number of operations required is no more than
$K\tau/2E_0+C(\epsilon_{\psi},\psi_0)\sqrt{K\tau}$, which is of order $K\tau$.

If the initial state $\ket{\psi}_{A_2B_2}$ is created with fidelity
$1-\epsilon_{\psi}$, then the fidelity of the final state resulting from the
communication process will satisfy
\begin{equation}
|\braket{\zeta}{\zeta_M}|^2 \ge 1-\epsilon_{\psi}-4M\sqrt{\epsilon}.
\end{equation}
Given any $\epsilon_0>0$, we may perform communication with fidelity
$1-\epsilon_0$ provided we create the initial state with fidelity
$1-\epsilon_0/2$, and $4M\sqrt{\epsilon}\le \epsilon_0/2$. The total
communication from Alice to Bob is $Mn_a\ge M\tau R'_\rightarrow$, and the total
communication from Bob to Alice is $Mn_b\ge M\tau R'_\leftarrow$. The total
number of operations does not exceed
$2M\tau+K\tau/2E_0+C(\epsilon_{\psi},\psi_0)\sqrt{K\tau}$, which does not exceed
$2M\tau+K\tau/2E_0+C(\epsilon_{\psi},\psi_0)\sqrt{K}\tau$. The communications per
operation from Alice to Bob and Bob to Alice are therefore at least
\begin{align}
& \frac{M\tau R'_\rightarrow}{2M\tau+\frac{K\tau}{2E_0}+C(\epsilon_0/2,\psi_0)
\sqrt{K}\tau}, \nn \\
& \frac{M\tau R'_\leftarrow}{2M\tau+\frac{K\tau}{2E_0}+C(\epsilon_0/2,\psi_0)
\sqrt{K}\tau}.
\end{align}
These rates will be at least $R_\rightarrow/2$ and $R_\leftarrow/2$ provided
\begin{align}
\label{achrat}
R'_\rightarrow-R_\rightarrow \ge \frac{R_\rightarrow K}{4ME_0} +
C(\epsilon_0/2,\psi_0) \frac {R_\rightarrow \sqrt{K}}{2M}, \nn \\
R'_\leftarrow-R_\leftarrow \ge \frac{R_\leftarrow K}{4ME_0} +
C(\epsilon_0/2,\psi_0) \frac {R_\leftarrow \sqrt{K}}{2M} .
\end{align}
Given any $\epsilon_0>0$, we may acheive the rate pair
$(R_\rightarrow/2,R_\leftarrow/2)$ with fidelity at least $1-\epsilon_0$ and
without initial entanglement by selecting the number of repetitions $M$ to be
sufficiently large so that inequalities \eqref{achrat} are satisfied. For this
value of $M$ we ensure that the fidelity $1-\epsilon_0$ is achieved by selecting
the block size $\tau$ to be sufficiently large so that the inequality
$4M\sqrt{\epsilon}\le \epsilon_0/2$ is satisfied.

Thus, for two-qubit unitary operations, if the rate pair $(R_\rightarrow,R_\leftarrow)$
is in the interior of $S_U^E$, the rate pair $(R_\rightarrow/2,R_\leftarrow/2)$
is achievable without entanglement assistance. As $R_\rightarrow+R_\leftarrow$ may
be taken to be arbitrarily close to $C_+^E$, we have the inequality for
two-qubit unitaries
\begin{equation}
C_+^E \le 2C_+.
\end{equation}
BHLS derive the inequality $C_+ \le E_U$, giving the ordering of the capacities
for two-qubit unitary operations
\begin{equation}
\label{order1}
C_+^E \le 2C_+ \le 2E_U.
\end{equation}

Similarly, $R_\rightarrow$ may be taken to be arbitrarily close to
$C_\rightarrow^E$, and $R_\leftarrow$ may be taken to be arbitrarily close to
$C_\leftarrow^E$, implying the further inequalities for two-qubit unitary operations
$C_\rightarrow^E \le 2C_\rightarrow$ and $C_\leftarrow^E \le 2C_\leftarrow$. For
two-qubit unitaries the communication capacities in each direction are
identical \cite{bennett}, so these inequalities may alternatively be expressed
as
\begin{equation}
C_{\rightarrow}^E=C_{\leftarrow}^E \le 2C_{\leftarrow}=2C_{\rightarrow}.
\end{equation}

Note that these results are crucially dependent on the fact that we have limited
the dimension of the ancilla that may be used. The number of operations required
to create the initial state scales no faster than $O(K\tau)$, and
$O(K\tau)/M\tau$ approaches zero in the limit $M\to\infty$. If the number of
operations required to create the initial state scaled superlinearly with
$\tau$, then it would not be possible to take this limit.

\section{Bidirectional ensembles}
\label{inccom}

A powerful means of deriving results for the case of communication in a
single direction is by considering the Holevo information of an ensemble of
states. When a joint state shared between Alice and Bob, $\ket{\psi_i}$, is
given with probability $p_i$, we denote this as the ensemble
${\cal E}=\{p_i,\ket{\psi_i}\}$. Similarly we may denote the ensemble of reduced
density matrices possessed by Bob by ${\sf E}=\{p_i,\rho_i\}$, where
$\rho_i={\rm Tr}_A(\ket{\psi_i}\bra{\psi_i})$. The index $i$ is chosen by Alice,
and may be used to communicate classical information. If a long sequence of the
states $\ket{\psi_i}$ is given with probabilities $p_i$, Alice may perform
communication equal to $\chi({\sf E})$ \cite{holevo,shuwes}.

BHLS show that the communication capacity in a single direction may be evaluated
by considering the Holevo information before and after applying the operation
$U$; that is, we may define
\begin{equation}
\label{holright}
\Delta\chi_U^\rightarrow = \sup_{\cal E} \left[ \chi({\rm Tr}_A U{\cal E})
-\chi({\rm Tr}_A{\cal E})\right].
\end{equation}
We take the convention that
\begin{align}
U{\cal E} &\equiv \{ p_i, U\ket{\Phi_i} \}, \\
{\rm Tr}_X{\cal E} &\equiv \{ p_i, {\rm Tr}_X(\ket{\Phi_i}) \},
\end{align}
where ${\rm Tr}_{X}(\ket{\psi}) \equiv {\rm Tr}_{X}(\ket{\psi}\bra{\psi})$. We
use the superscript ``$\rightarrow$'' to indicate that this is the difference in the
Holevo information for communication from Alice to Bob. We may similarly define
\begin{equation}
\Delta\chi_U^\leftarrow = \sup_{\cal E} \left[ \chi({\rm Tr}_B U{\cal E})
-\chi({\rm Tr}_B{\cal E})\right],
\end{equation}
for communication from Bob to Alice. It is shown in Ref.\ \cite{bennett} that
$C_\rightarrow^E=\Delta\chi_U^\rightarrow$ and $C_\leftarrow^E=\Delta
\chi_U^\leftarrow$.

Here we apply similar ideas to the case of bidirectional communication. In this
case, let $i$ be the message encoded by Alice and $j$ be the message encoded by
Bob. We may then define a bidirectional ensemble by
\begin{equation}
{\cal E}=\{p_i,q_j,\ket{\psi_{ij}}\}.
\end{equation}
In general, Bob has a reduced density matrix that depends on $i$ and $j$, which
he may use to obtain information about $i$. Similarly, Alice has a reduced
density matrix dependent on $i$ and $j$, and may perform measurements to gain
information about $j$.

In order to apply the same ideas as for the case of communication in a single
direction, it is necessary to consider the bidirectional communication that may
be performed using this ensemble. Here we give upper and lower bounds, and we
will show in Sec.\ \ref{bidir} that these bounds coincide for a particular type
of ensemble.

In order to obtain an upper bound, we may assume that Alice and Bob use
information about the other's message in their encoding. If Alice knows in
advance the values of $j$ that Bob will be using, then she may perform block
coding over those ensembles with the same value of $j$. The information that
Alice may communicate to Bob for each value of $j$ is given by
\begin{equation}
S\left( \sum_i p_i \rho_{ij}^B\right)
- \sum_i p_i S(\rho_{ij}^B),
\end{equation}
where $\rho_{ij}^B={\rm Tr}_A\ket{\psi_{ij}}$.
Averaging over the values of $j$, the average communication is equal to  
\begin{equation}
\label{right}
\chi_{\rm up}^{\rightarrow}({\cal E}) = \sum_j q_j \! \left[ S\left( \sum_i p_i
\rho_{ij}^B\right) - \sum_i p_i S(\rho_{ij}^B)\right].
\end{equation}
When Alice does not have advance information about the values of $j$, she will
not be able to perform the coding in this way. Nevertheless, it is obvious that
it will not be possible to perform more communication without prior information
about the values of $j$. Therefore $\chi_{\rm up}^{\rightarrow}({\cal E})$ gives
an upper limit to the communication that may be performed from Alice to Bob
using this ensemble. In exactly the same way, we see that an upper limit to the
communication possible from Bob to Alice is given by
\begin{equation}
\label{left}
\chi_{\rm up}^{\leftarrow}({\cal E}) = \sum_i p_i \! \left[ S\left( \sum_j q_j
\rho_{ij}^A\right) - \sum_j q_j S(\rho_{ij}^A)\right],
\end{equation}
where $\rho_{ij}^A={\rm Tr}_B\ket{\psi_{ij}}$.

We may obtain a lower bound by assuming that Alice and Bob code independently
of the messages being encoded by the other party. If Alice
chooses message $i$, and has no information about $j$, she may take Bob's
density matrix to be $\rho_i^B = \sum_j q_j \rho_{ij}^B$.

In order to improve upon this, Bob may apply some $j$-dependent completely
positive map ${\cal T}_j^B$ to his portion of the state. The idea behind using
this map is to reduce the $j$ dependence of Bob's reduced density matrix, so the
averaged density matrices $\rho_i^B$ are more distinguishable for different
values of $i$. With this operation, Alice may take Bob's density matrix to be
\begin{equation}
\label{rho}
\rho_i^B = \sum_j q_j {\cal T}_j^B ( \rho_{ij}^B ) .
\end{equation}
Via coding over multiple
ensembles, Alice may perform communication to Bob equal to the Holevo
information of the ensemble $\{p_i,\rho_i^B\}$. We now take the supremum of this
over the maps $\{{\cal T}_j^B\}$, and define
\begin{equation}
\chi_{\rm lo}^{\rightarrow}({\cal E}) = \sup_{\{{\cal T}_j^B\}} \chi
(\{ p_i, \rho_i^B \}),
\end{equation}
where $\rho_i^B$ is defined as in Eq.\ \eqref{rho}. This is a lower limit to
the communication that may be performed from Alice to Bob. Similarly a lower
limit to the communication that may be performed from Bob to Alice is
\begin{equation}
\chi_{\rm lo}^{\leftarrow}({\cal E}) = \sup_{\{{\cal T}_i^A\}} \chi
(\{ q_j, \rho_j^A \}),
\end{equation}
where
\begin{equation}
\rho_j^A = \sum_i p_i {\cal T}_i^A (\rho_{ij}^A ),
\end{equation}
and ${\cal T}_i^A$ is a completely positive map applied by Alice.

We will also denote the exact values of the maximum communication that it is
possible to communicate from Alice to Bob and from Bob to Alice by
$\chi^{\rightarrow}({\cal E})$ and $\chi^{\leftarrow}({\cal E})$, respectively.
From the definitions, it is evident that
\begin{equation}
\label{ineqs}
\chi_{\rm lo}^{\rightarrow}({\cal E}) \le \chi^{\rightarrow}({\cal E}) \le
\chi_{\rm up}^{\rightarrow}({\cal E}), ~~~
\chi_{\rm lo}^{\leftarrow}({\cal E}) \le \chi^{\leftarrow}({\cal E}) \le
\chi_{\rm up}^{\leftarrow}({\cal E}).
\end{equation}
We may also consider the maximum bidirectional communication that it is possible
to perform with the ensemble,
$\chi^{\leftrightarrow}({\cal E}) = \chi^{\rightarrow}({\cal E}) +
\chi^{\leftarrow}({\cal E})$. This quantity has the upper and lower bounds
\begin{equation}
\chi_{\rm lo}^{\leftrightarrow}({\cal E}) =
 \chi_{\rm lo}^{\rightarrow}({\cal E}) + \chi_{\rm lo}^{\leftarrow}({\cal E}),
~~~ \chi_{\rm up}^{\leftrightarrow}({\cal E}) =
 \chi_{\rm up}^{\rightarrow}({\cal E}) + \chi_{\rm up}^{\leftarrow}({\cal E}).
\end{equation}

Analogous to the case for communication in a single direction, we may define
the maximum that the classical communication capacity of an ensemble may be
increased,
\begin{equation}
\Delta\chi_U^{\leftrightarrow} = \sup_{\cal E} \left[ \chi^{\leftrightarrow}
(U{\cal E})-\chi^{\leftrightarrow}({\cal E})\right] .
\end{equation}
This quantity is analogous to the unidirectional quantity
$\Delta\chi_U^{\rightarrow}$ introduced by BHLS. However, in contrast to the
unidirectional case, it has not been proven that
$C_+^E = \Delta \chi_U^\leftrightarrow$.

\section{Ensembles for entanglement-assisted communication}
\label{ensemb}
In this section we show how to obtain initial ensembles for two-qubit operations
such that the communication possible in one or both directions may be increased
by $E_U$. We first consider the simpler case of communication in a single
direction, then extend this to the bidirectional case.

\subsection{Communication in a single direction}
\label{unidir}
We consider the case of $\Delta\chi_U^\rightarrow$. As all two-qubit operations
are symmetric, the results for this case also apply to
$\Delta\chi_U^\leftarrow$. Note that the second term on the right-hand side of
Eq.\ \eqref{holevo} is the average of the entanglement of the coding states.
Therefore, if each of the states of the initial ensemble are chosen such that
the entanglement of these states is decreased by the maximum $E_U$ by
operation $U$, then the second term in Eq.\ \eqref{holevo} will be decreased by
$E_U$ by the operation. If the first term is constant, then the total increase
in the Holevo information will be $E_U$.

In order to obtain such an ensemble, let us start with an initial state
$\ket{\Psi}$ such that the entanglement is decreased by the maximum $E_U$ via
operation $U$. (Recall that $E_U=E_U^-$ for two-qubit operations.) We then
obtain the coding states via a set of operations $\{ V_i^A \otimes V_i^B \}$,
where $V_i^A$ and $V_i^B$ are local operations on Alice's and Bob's sides,
respectively. We use the notation where (1) indicates Alice's ancilla, (2)
indicates Alice's qubit upon which $U$ acts, (3) indicates Bob's qubit upon
which $U$ acts, and (4) indicates Bob's ancilla. The ancillas may be qubits or
arbitrary dimensional qudits. In addition, we use $A$ to indicate
Alice's entire system (1 and 2), and we use $B$ to indicate Bob's entire system
(3 and 4). In this section we indicate the subsystem with superscripts (rather
than subscripts) to avoid notational conflicts.

The operations $V_i\equiv V_i^A\otimes V_i^B$ must satisfy two criteria.

(1) $V_i$ commutes with $U$. This means that each state $V_i \ket{\Psi}$ will
have its entanglement decreased by $E_U$ for operation $U$.

(2) For all states $\ket{\phi}$,
\begin{equation}
\label{crit2}
\sum_i p_i {\rm Tr}_A (V_i \ket{\phi}) = \tfrac 12 \openone\otimes \rho^{(4)},
\end{equation}
where $\rho^{(4)}={\rm Tr}_{A3}(\ket{\phi})$ is the reduced density matrix for
Bob's ancilla (4).

As the operation $U$ does not act on the ancilla, $\rho^{(4)}$ will be unchanged
under $U$. That is,
\begin{equation}
\sum_i p_i{\rm Tr}_A (V_i\ket{\Psi})=\sum_i p_i{\rm Tr}_A (V_iU\ket{\Psi}) 
=\tfrac 12 \openone \otimes \rho^{(4)}.
\end{equation}
This means that the first term on the right side of Eq.\ \eqref{holevo} will be
unchanged by the operation $U$, and therefore that the total one-way
communication possible is $E_U$.

For $U$ in the form of Eq.\ \eqref{simpleU}, the above conditions are satisfied
for the operators $V_i=\sigma_i^{(2)}\otimes\sigma_i^{(3)}$, for
$i\in\{0,1,2,3\}$. Here $\sigma_i$ are the Pauli matrices for
$i\in\{1,2,3\}$, $\sigma_0$ is the identity, and the superscripts indicate the
qubits upon which the operators act. It is easy to see that these
operators commute with $U$ in the form of Eq.\ \eqref{simpleU}. To show the second
property, let us express the reduced density matrix at Bob's side,
$\rho^B={\rm Tr}_A (\ket{\Psi})$, in the form
\begin{equation}
\rho^B = \tfrac 12 \sum_{k=0}^3 \sigma_k^{(3)} \otimes \rho_k^{(4)}.
\end{equation}
It is easily seen that, taking $p_i=1/4$, we obtain
\begin{equation}
\sum_i p_i {\rm Tr}_A (V_i\ket{\Psi}) = \sum_i p_i \sigma_i^{(3)} \rho^B
\sigma_i^{(3)} = \tfrac 12 \sigma_0^{(3)} \otimes \rho_0^{(4)}.
\end{equation}

Since the trace over qubit (3) gives $\rho^{(4)}$, and the trace of
each of $\sigma_1$, $\sigma_2$, and $\sigma_3$ is equal to zero, we must have
$\rho_0^{(4)} = \rho^{(4)}$. We therefore obtain Eq.\ \eqref{crit2},
satisfying the second condition for the operations $V_i$. Thus we find that, by
coding with equal probabilities $p_i=1/4$ each of the four states $\sigma_i^{(2)}
\sigma_i^{(3)}\ket{\Psi}$, it is possible to increase the Holevo information by
$E_U$. Thus we have proven the inequality $\Delta\chi_U^\rightarrow\ge E_U$.
Using the equality $C_\rightarrow^E=\Delta \chi_U^\rightarrow$ \cite{bennett},
we have the inequality $C_{\rightarrow}^E\ge E_U$.

\subsection{Bidirectional communication}
\label{bidir}
Next we apply a similar coding protocol to the more complicated case of
bidirectional communication. We start with a state $\ket\Psi$ such that
$U$ decreases the entanglement by the maximum $E_U$. Alice encodes via the set
of four operators $\{ \sigma_i^{(2)}\sigma_i^{(3)} \}$, and Bob encodes via the
four operators $\{\sigma_j^{(2)}\sigma_j^{(3)}\}$. Alice's and Bob's
operators commute, so the order in which these operators are applied is
irrelevant. We therefore have a total ensemble of 16 states
\begin{equation}
\label{bidirens}
{\cal E} = \{ p_i, q_j, \sigma_i^{(2)} \sigma_i^{(3)} \sigma_j^{(2)}
\sigma_j^{(3)} \ket\Psi \},
\end{equation}
where we take $p_i=q_j=1/4$. Note that each of the operators applied by Alice
and Bob commutes with $U$, for $U$ in the form of Eq.\ \eqref{simpleU}. Therefore, for
both the initial and final ensembles, the members of the ensemble are of the
form $\ket{\psi_{ij}}=\sigma_i^{(2)}\sigma_i^{(3)}\sigma_j^{(2)}\sigma_j^{(3)}
\ket{\phi}$, where $\ket{\phi}=\ket{\Psi}$ for the initial ensemble and
$\ket{\phi}=U\ket{\Psi}$ for the final ensemble. This simple form allows us to
determine the exact bidirectional communication that it is possible to perform.

We now take the completely positive maps ${\cal T}_i^A$ and ${\cal T}_j^B$
to be the simple unitary operations
\begin{equation}
{\cal T}_i^A(\rho) = \sigma_i^{(2)} \rho \sigma_i^{(2)}, ~~~~~
{\cal T}_j^B(\rho) = \sigma_j^{(3)} \rho \sigma_j^{(3)}.
\end{equation}
This gives
\begin{align}
\rho_i^B &= \sum_j q_j {\cal T}_j^B [\sigma_i^{(3)}\sigma_j^{(3)}
{\rm Tr}_A(\ket{\phi})\sigma_j^{(3)}\sigma_i^{(3)} ] \nn \\
&= \sum_j q_j [\sigma_i^{(3)}{\rm Tr}_A(\ket{\phi})\sigma_i^{(3)} ] \nn \\
&= \sigma_i^{(3)}{\rm Tr}_A(\ket{\phi})\sigma_i^{(3)} ,
\end{align}
and similarly
\begin{align}
\rho_j^A &= \sum_i p_i {\cal T}_i^A [\sigma_i^{(2)}\sigma_j^{(2)}
{\rm Tr}_B(\ket{\phi})\sigma_j^{(2)}\sigma_i^{(2)} ] \nn \\
&= \sigma_j^{(2)}{\rm Tr}_B(\ket{\phi})\sigma_j^{(2)} .
\end{align}
We then obtain the Holevo information of the ensembles ${\sf E}^B=
\{p_i,\rho_i^B\}$ and ${\sf E}^A=\{q_j,\rho_j^A\}$ as
\begin{align}
\label{first}
\chi({\sf E}^B)&=S(\tfrac 12 \openone\otimes\rho^{(4)})-
S[{\rm Tr}_A(\ket{\phi})]=\chi_{\rm up}^{\rightarrow}({\cal E}), \\
\label{second}
\chi({\sf E}^A)&=S(\tfrac 12\rho^{(1)}\otimes\openone)-
S[{\rm Tr}_B(\ket{\phi})]=\chi_{\rm up}^{\leftarrow}({\cal E}).
\end{align}
Here we have used
\begin{align}
\sum_i {\rm Tr}_A \big( \sigma_i^{(2)}\sigma_i^{(3)}\sigma_j^{(2)}\sigma_j^{(3)}
\ket{\phi}\big) &\propto  \openone \otimes \rho^{(4)}, \\
\sum_j {\rm Tr}_B \big( \sigma_i^{(2)}\sigma_i^{(3)}\sigma_j^{(2)}\sigma_j^{(3)}
\ket{\phi}\big) &\propto  \rho^{(1)} \otimes \openone,
\end{align}
where $\rho^{(1)}={\rm Tr}_{2B} (\ket{\phi})$ is the reduced density matrix for
Alice's ancilla.

As we have taken the lower limits $\chi_{\rm lo}^\rightarrow({\cal E})$ and
$\chi_{\rm lo}^\leftarrow({\cal E})$ to be the supremums over completely
positive maps of $\chi({\sf E}^B)$ and $\chi({\sf E}^A)$, and we have
inequalities \eqref{ineqs}, we must have
\begin{equation}
\chi_{\rm lo}^\rightarrow({\cal E})=\chi_{\rm up}^\rightarrow({\cal E})=
\chi^\rightarrow({\cal E}), ~~~
\chi_{\rm lo}^\leftarrow({\cal E})=\chi_{\rm up}^\leftarrow({\cal E})=
\chi^\leftarrow({\cal E}).
\end{equation}
This implies that $\chi_{\rm lo}^\leftrightarrow({\cal E})=
\chi_{\rm up}^\leftrightarrow({\cal E})=\chi^\leftrightarrow({\cal E})$.

Therefore, for this specific type of bidirectional ensemble, the
upper and lower limits on the bidirectional communication that it is possible
to perform coincide, and we have an explicit expression for the
bidirectional communication that may be performed. To understand the reason for
this, Bob performs local operations that remove the $j$ dependence of his part
of the shared entangled state. This allows Alice to perform coding without
regard for the value of $j$. Similarly, Bob may perform coding without regard
for the value of $i$.

Now note that, under the action of the operation $U$, the first terms in
Eqs.\ \eqref{first} and \eqref{second} are constant (because the operation
does not act upon the ancillas). Also the entanglement
$S[{\rm Tr}_A(\ket{\phi})]=S[{\rm Tr}_B(\ket{\phi})]$ decreases by $E_U$.
Therefore the total increase in the bidirectional communication possible is
equal to $2E_U$.

As we have given an explicit scheme that increases the bidirectional
communication that it is possible to perform by $2E_U$, we have proven the
inequality for two-qubit operations $2E_U \le \Delta \chi_U^\leftrightarrow$.
Using the other inequalities derived in the preceding sections, we have the
sequence of inequalities
$C_+^E\le 2C_+\le 2E_U\le \Delta\chi_U^\leftrightarrow$.

For the cases of the {\sc cnot} and {\sc swap} operations, these inequalities are
equalities. A potential way of proving equalities in the general case is to
show that ensembles of the form of Eq.\ \eqref{bidirens} can be prepared using average
bidirectional communication of $\chi^\leftrightarrow({\cal E})$ per ensemble.
Such preparation is efficient in the sense that the communication required to
prepare the ensembles is the same as what may be performed using the ensembles.

In the unidirectional case, it is known that ensembles can be prepared
efficiently \cite{Shor,berry}. BHLS apply this result to show that, given an
ensemble ${\cal E}$ such that the Holevo information is increased by
$\Delta \chi = \chi({\rm Tr}_A U{\cal E})-\chi({\rm Tr}_A{\cal E})$, the
entanglement-assisted communication capacity in a single direction is at least
$\Delta \chi$. This proof is based upon a communication scheme where initial
communication is obtained by some finite number of operations, and used to
create initial ensembles. The Holevo information of these ensembles is increased
via application of the operation $U$ and used to obtain further communication.
This communication may be used to obtain further initial ensembles, and the
process is repeated. This process may be repeated a large number of times, so
that the average communication per operation is equal to $\Delta\chi$.

It is clear that a similar communication scheme could be applied in the
bidirectional case, if it were possible to efficiently create initial ensembles.
Not all bidirectional ensembles can be prepared efficiently. For example,
consider the ensemble ${\cal E}=\{ 1/2, 1/2, \ket{\psi_{ij}}\}$ where
$\ket{\psi_{00}}=\ket{\psi_{01}}=\ket{\psi_{10}}=\ket{00}$ and
$\ket{\psi_{11}}=\ket{11}$. It is easily seen that this ensemble
can only be used to perform a total bidirectional communication of 1 bit,
but it requires 2 bits of communication to create.

However, there are some bidirectional ensembles that can be efficiently
prepared, and it is plausible that ensembles of the form \eqref{bidirens} can be
efficiently prepared. Even though it is not known how to efficiently create
these ensembles, it is possible to efficiently create ensembles with the same
reduced density matrices (see Appendix \ref{appB}). Because the bidirectional
communication that can be performed using the ensemble of Eq.\ \eqref{bidirens} is
increased by $2E_U$ via the operation $U$, if it is possible to efficiently
create ensembles of this type, it should be possible to apply a communication
scheme similar to that used by BHLS such that the average communication is
$2E_U$. Such a result would imply that $C_+^E \ge 2E_U$, and therefore
$C_+^E=2C_+=2E_U$.

\section{Conclusions}
\label{conclude}
By considering a quantum superposition of classical messages, we have shown that
the capacities for general nonlocal unitary operations satisfy the inequality $C_+^E\le E_U+E_U^-$. For the
case of two-qubit unitary operations, we have the further inequalities
\begin{equation}
C_+^E\le 2C_+\le 2E_U.
\end{equation}
In order to show these inequalities, we have demonstrated that it is possible to
put an upper limit on the dimension of the ancilla required for the
communication process. This upper limit also simplifies some of the derivations
given by BHLS.

We have given an explicit scheme for finding ensembles of states such that the
Holevo information may be increased by $E_U$ for a two-qubit unitary operation $U$.
Together with the result given by BHLS $C_\rightarrow^E=\Delta
\chi_U^\rightarrow$, this implies that $E_U\le C_\rightarrow^E$.

We have also introduced the concept of bidirectional ensembles, and given
upper and lower bounds for the communication that may be performed
using these ensembles. For two-qubit unitary operations, we have shown that the
bidirectional communication that an ensemble can achieve may be
increased by at least $2E_U$; that is, we have the series of inequalities
for two-qubit unitary operations
\begin{equation}
\label{inequals}
C_+^E\le 2C_+\le 2E_U\le \Delta\chi_U^\leftrightarrow.
\end{equation}
A promising way of proving equalities is by developing an efficient scheme for
remote state preparation of bidirectional ensembles. This problem is a possible
direction for future research. 

\acknowledgments
We thank Aram Harrow for providing the example illustrating that not all
bidirectional ensembles can be prepared efficiently.
This project has been supported by an Australian Research Council Large Grant.
We also acknowledge valuable discussions with T.\ J.\ Johnson and J.\ Daboul.

\appendix
\section{Communication rates}
\label{def}
We may express the set of achievable rate pairs without entanglement assistance
by
\begin{align}
S_U &= \big\{ (R_{\rightarrow},R_{\leftarrow}) | \forall \epsilon>0, \exists t
\in {\mathbb Z}^+,\exists n_a \ge tR_{\rightarrow}, \nn \\ & \exists
n_b \ge t R_{\leftarrow},\exists \big\{ V_A^{(j)} \otimes V_B^{(j)}
\big\}_{j=0}^t, \exists d_1,d_2 \in {\mathbb Z}^+ \nn \\
& {\rm s.t.} ~~~ \forall x \in \{ 0,1 \}^{n_a}, \forall y \in \{ 0,1 \}^{n_b},
\nn \\ & F\left( \ket{y}_{A_1} \ket{x}_{B_1}, {\rm Tr}_{A_2B_2}
\ket{\eta_{xy}}_{AB}\bra{\eta_{xy}} \right) \ge 1-\epsilon, \nn \\
& {\rm for} ~~~ \ket{\eta_{xy}}_{AB} = (V_A^{(t)} \otimes V_B^{(t)}) U
\cdots  \nn \\  & \cdots U (V_A^{(0)} \otimes V_B^{(0)}) \ket{x}_{A_1}
\ket{y}_{B_1} \ket{0_{d_1}}\ket{0_{d_2}} \big\} ,
\end{align}
where $\ket{0_{d_1}}$ and
$\ket{0_{d_2}}$ are pure states of ancilla subsystems possessed by Alice
and Bob with dimensions $d_1$ and $d_2$, respectively. The set of achievable
rate pairs with entanglement assistance is defined by
\begin{align}
S_U^E &= \big\{ (R_{\rightarrow},R_{\leftarrow}) | \forall \epsilon>0,
\exists t \in {\mathbb Z}^+,\exists n_a \ge tR_{\rightarrow}, \nn \\
& \exists n_b \ge t R_{\leftarrow},\exists \big\{ V_A^{(j)} \otimes V_B^{(j)}
\big\}_{j=0}^t, \exists d_1,d_2,d_3 \in {\mathbb Z}^+ \nn \\
& {\rm s.t.} ~~~ \forall x \in \{ 0,1 \}^{n_a}, \forall y \in \{ 0,1 \}^{n_b},
\nn \\ & F\left( \ket{y}_{A_1} \ket{x}_{B_1}, {\rm Tr}_{A_2B_2}
\ket{\eta_{xy}}_{AB}\bra{\eta_{xy}} \right) \ge 1-\epsilon, \nn \\
& {\rm for} ~~~ \ket{\eta_{xy}}_{AB} = (V_A^{(t)} \otimes V_B^{(t)}) U
\cdots  \nn \\  & \cdots U (V_A^{(0)} \otimes V_B^{(0)}) \ket{x}_{A_1}
\ket{y}_{B_1} \ket{0_{d_1}}\ket{0_{d_2}}\ket{\Phi_{d_3}}
\big\} ,
\end{align}
where $\ket{\Phi_{d_3}}$ is a maximally entangled state with each subsystem of
dimension $d_3$.

\section{Ensemble creation}
\label{appB}
An ensemble for which the reduced density matrices possessed by Alice and Bob
are the same as for the ensemble of Eq.\ \eqref{bidirens} is given by
\begin{align}
\label{densens}
{\cal E}&=\{p_i,q_j,\sigma_i^{(2)}\sigma_j^{(2)}{\rm Tr}_B(\ket{\phi})
\sigma_j^{(2)}\sigma_i^{(2)} \nn \\ &\otimes \sigma_i^{(3)}\sigma_j^{(3)}
{\rm Tr}_A(\ket{\phi})\sigma_j^{(3)}\sigma_i^{(3)} \},
\end{align}
where $p_i=q_j=1/4$.

In order to efficiently prepare this ensemble, Alice and Bob efficiently prepare
the two unidirectional ensembles $\{ p_i, \sigma_i^{(3)}{\rm Tr}_A(\ket{\phi})
\sigma_i^{(3)} \}$ and $\{ q_j, \sigma_j^{(2)}{\rm Tr}_B(\ket{\phi})
\sigma_j^{(2)} \}$. Here $i$ is the index chosen by Alice and $j$ is the
index chosen by Bob. These two unidirectional ensembles may be treated as one
bidirectional ensemble
\begin{equation}
\{p_i,q_j,\sigma_j^{(2)}{\rm Tr}_B(\ket{\phi})\sigma_j^{(2)} \sigma_i^{(3)}
{\rm Tr}_A(\ket{\phi})\sigma_i^{(3)} \}.
\end{equation}
In order to obtain the ensemble given in Eq.\ \eqref{densens}, Alice
performs the local operation $\sigma_i^{(2)}$, and Bob performs the local
operation $\sigma_j^{(3)}$.

\end{document}